\documentclass[11pt,a4paper]{article}
\usepackage{jheppub}
\usepackage{subfig}

\def\Id{\mathbb I}

\def\half{\textstyle{1\over2}}

\newcommand{\del}{\ensuremath{\partial}}
\newcommand{\ba}{\begin{eqnarray}}
\newcommand{\ea}{\end{eqnarray}}
\newcommand{\Tr}{\textrm{Tr}}

\title{Dynamical simulations of electroweak baryogenesis with fermions}

\author[a]{Paul M. Saffin,}
\author[b]{Anders Tranberg}

\affiliation[a]{School of Physics and Astronomy, University Park, University of Nottingham,\\ Nottingham NG7 2RD, United Kingdom}
\affiliation[b]{Niels Bohr International Academy and Discovery Center, Niels Bohr Institute,\\  Blegdamsvej 17, 2100 Copenhagen, Denmark}

\emailAdd{paul.saffin@nottingham.ac.uk}
\emailAdd{anders.tranberg@nbi.dk}

\abstract{We perform real-time numerical lattice simulations of a one-family version of the Standard Model. We model the quantum fermions using the ensemble method and treat the bosonic scalar and non-abelian gauge fields classically. Our main interest is electroweak baryogenesis, and we test the approach by considering Standard Model baryon number violation through the chiral anomaly, a truly quantum phenomenon. We find that the method correctly reproduces the anomaly, and perform the first dynamical simulations of electroweak baryon number violation including fermions.}


\begin{document}

\maketitle

\section{Introduction}

Since the conditions for generating a matter-antimatter asymmetry were laid out by Sakharov \cite{Sakharov:1967dj}, many scenarios for baryogenesis have been  proposed, based on high-energy processes in the early Universe (for a review, see for instance \cite{Dine:2003ax}). Here we report on progress of a technique that can be used to examine various models of electroweak-scale baryogenesis \cite{Kuzmin:1985mm,Rubakov:1996vz,Cohen:1993nk}, whose common feature is that the source of baryon number violation is the electroweak anomaly \cite{'tHooft:1976up}.  

All the key symmetry violations (CP, C and baryon number) occur in the Standard Model, and the requirement of departure from thermal equilibrium is determined by early-Universe dynamics and the electroweak symmetry breaking transition. In the Standard Model (SM), CP violation is provided by the CKM matrix of quark mixings \cite{Cabibbo:1963yz,Kobayashi:1973fv} and possibly by the PMNS matrix of neutrino mixings \cite{Pontecorvo:1957qd,Maki:1962mu}, although this is less well measured experimentally. Beyond the SM, CP-violation can be provided for instance through Majorana masses, multiple Higgs field with complex coupling and/or higher-dimensional effective terms.

In electroweak baryogenesis the baryon asymmetry is expected to occur at the electroweak scale of $\sim$100GeV, at which time the Hubble expansion rate is around $10^{-5}$eV - very slow by the standards of microscopic electroweak processes. Given that a thermal electroweak phase transition in the Standard Model cannot be first order for current Higgs-mass bounds \cite{Kajantie:1995kf,Csikor:1998eu} , we are left with the problem of finding dynamics that can cause sufficient departure from thermal equilibrium. Such a possibility is raised by models inspired by preheating at the end of inflation, where the phase transition is non-thermal \cite{GarciaBellido:1999sv,Krauss:1999ng,Copeland:2001qw,Rajantie:2000nj,Copeland:2002ku,GarciaBellido:2002aj,Smit:2002yg,GarciaBellido:2003wd}, or in extensions of the Standard Model with a more complicated scalar sector, where the phase transition can be strongly first order \cite{Fromme:2006cm}.

Whilst it is nowadays a relatively straightforward task to simulate the classical dynamics of the bosonic degrees of freedom of the Standard Model and minimal extensions, using techniques from lattice gauge theory, the fermions are more problematic. And given that all the CP violation of the Standard Model is in the fermion sector, and the baryon number itself is a fermion quantum number these cannot merely be ignored. One approach is to "integrate out" the fermions, and treat their dynamics as new terms in an effective action. This approach has been taken in a number of studies \cite{Ambjorn:1990pu,Tranberg:2003gi,Tranberg:2009de,Tranberg:2010af} (see also \cite{Hernandez:2008db,GarciaRecio:2009zp,Brauner:2011vb}).

Here we present a numerical method that evolves the fermions along with the bosons and, when extended to three families of fermions, can include CP violation from the fermion sector directly.

The technique that we use to simulate the dynamical fermions was first laid out in \cite{Borsanyi:2008eu}, with a sample study of the effect of fermions on the evolution of oscillons in 2+1 dimensions. The idea is to model the quantum averages $\langle\;|...|\;\rangle$ by ensemble averages $\langle...\rangle_e$, and so by simulating an ensemble of fermion realizations, one is effectively sampling the evolution of the whole set of fermion mode functions. The utility of this is that the number of mode functions varies as $n_x^3$ in three-dimensions (here $n_x$ is the number of lattice sites in one direction), which is prohibitively expensive numerically, whereas the number of modes required to give a reasonable sample is expected to vary as $n_x$ \cite{Borsanyi:2008eu} (see also the 3+1D simulations in a scalar-fermion theory in \cite{bergesgelfand}). 

One check that the method is viable for reproducing the quantum anomaly has already been performed in an axial Yang-Mills-Higgs-fermion system in 1+1 dimensions \cite{Saffin:2011kc}, where the baryon-number violating processes where directly simulated. The results of the ensemble method were in complete agreement with the evolution using the full set of mode functions \cite{Aarts:1998td,Aarts:1999zn}. While there is no numerical gain in 1+1 dimensions by using the ensemble method, the results of \cite{Saffin:2011kc} gives confidence in the technique, and provided motivation for the present 3+1 study, which would not be practical using the full set of mode functions, certainly for lattices of the size used in \cite{Tranberg:2009de}.

In this paper we use the fermion ensemble method to simulate the electroweak quantum anomaly directly. In this first study we shall be using only a single family of Standard Model fermions \footnote{We include a right-handed neutrino in our definition of the Standard Model.} which is enough to establish the viability of the method by direct reproduction of the anomaly. We shall also be omitting the U(1) hypercharge and SU(3) colour of the Standard Model, as this does not affect the anomaly process.

The structure of the paper is as follows: We start in section \ref{sec:continuum} with a presentation of the continuum model that we aim to simulate and in section \ref{sec:modelling} discuss how we treat the fermions and bosons numerically. In an equilibrium environment, the key baryon number violating process is the sphaleron transition. In section \ref{sec:sphalerons} we test the real-time fermions in the background of handmade sphalerons. We then describe how to do full boson-fermion dynamics in the context of a fast electroweak transition in \ref{sec:tacres}, where we demonstrate the approach with sample simulations. We conclude in section \ref{sec:conclusion}. In order to keep a flow to the paper we have included a number of the technical details in the appendices, including a list of conventions in App. \ref{app:conventions}, the lattice implementation in App. \ref{app:latticeAction}, the fermion initial conditons in App. \ref{app:ferminit} and the details of the handmade sphalerons in App. \ref{app:HMsphalerons}

\section{The SU(2)-Higgs model with chiral fermions in 3+1 dimensions }
\label{sec:continuum}

Our field content is one family of the Standard Model (extended to include a right-handed neutrino), but without U(1) hypercharge, SU(3) gluons and without colour degrees of freedom. Therefore we have a complex Higgs doublet $\phi$; SU(2) gauge fields $W^a_\mu$; a left-handed SU(2)-doublet quark field $q_L=\left(u_L,d_L\right)$; two right-handed SU(2)-singlet quark fields $u_R$, $d_R$; a left-handed SU(2)-doublet lepton field $l_L=\left(\nu_L,e_L\right)$; and two right-handed SU(2)-singlet lepton fields $e_R$, $\nu_R$. The continuum action is then written as
\ba
\label{eq:contAcReduced}
S=S_H+S_W+S_{F}+S_{Y},
\ea
where the set of different components is given by
\ba
S_H&=&-\int\;d^4x\;\left[D_\mu\phi^\dagger D^\mu\phi+\lambda(\phi^\dagger\phi-v^2/2)^2\right],\\
S_W&=&-\int\;d^4x\;\frac{1}{4}W^a_{\mu\nu}W^{a,\mu\nu},\\
S_F&=&-\int\; d^4x\;\left[ \bar{q}_L\gamma^\mu D_\mu q_L
                          +\bar{u}_R\gamma^\mu D_\mu u_R+\bar{d}_R\gamma^\mu D_\mu d_R\right.\nonumber\\
                          &~&\qquad\qquad\left.+\bar{l}_L\gamma^\mu D_\mu l_L
                          +\bar{\nu}_R\gamma^\mu D_\mu \nu_R+\bar{e}_R\gamma^\mu D_\mu e_R\right],\\
S_{Y}&=&-\int\; d^4x\;\left[ 
          G^u\bar{q}_L\phi u_R+G^d\bar{q}_L\phi d_R+G^e\bar{l}_L\phi e_R+G^\nu\bar{l}_L\phi \nu_R\right.\\\nonumber
       &~&\qquad\qquad+\hat G^u\bar{q}_L\tilde\phi u_R+\hat G^d\bar{q}_L\tilde\phi d_R+\hat G^e\bar{l}_L\tilde\phi e_R+\hat G^\nu\bar{l}_L\tilde\phi\nu_R\\\nonumber
       &~&\left.\qquad\qquad+h.c.\right].
\ea
The absence of a U(1) hypercharge means that one is allowed more Yukawa terms than the equivalent Standard Model action, but we simply set to zero those that would not be allowed had we included the Standard Model U(1).

The charges of the fields are determined by the covariant derivatives, which are
\ba
\label{eq:covDer_phi}
D_\mu\phi&=&\left(\del_\mu-\frac{ig}{2}\sigma^aW^a_\mu\right)\phi,\\
D_\mu q_L&=&\left(\partial_\mu-\frac{ig}{2}\sigma^aW^a_\mu\right)q_L,\quad D_\mu u_R=\partial_\mu u_R,\quad D_\mu d_R=\partial_\mu d_R,\\
D_\mu l_L&=&\left(\partial_\mu-\frac{ig}{2}\sigma^aW^a_\mu\right)l_L\quad D_\mu e_R=\partial_\mu e_R,\quad D_\mu \nu_R=\partial_\mu \nu_R,
\ea
and the SU(2) field-strength is defined by
\ba
\left[D_\mu,D_\nu\right]\phi&=&-\frac{ig}{2}\sigma^aW^a_{\mu\nu}\phi.
\ea
As well as the conserved current following from the presence of the gauge symmetry, there is a baryon current and a lepton current coming from the global symmetries $q\rightarrow\exp(i\alpha\gamma^5)q$ and $l\rightarrow\exp(i\tilde\alpha\gamma^5)l$,
\ba
j^\mu_{(b)}&=&i\left[\bar q_L\gamma^\mu q_L+\bar u_R\gamma^\mu     u_R+\bar d_R\gamma^\mu d_R\right]=i\bar q\gamma^\mu q,\\
j^\mu_{(l)}&=&i\left[\bar l_L\gamma^\mu l_L+\bar \nu_R\gamma^\mu \nu_R+\bar e_R\gamma^\mu e_R\right]=i\bar l\gamma^\mu l.
\ea
Classically, these currents are conserved. Quantum mechanically however, one finds that \cite{Fujikawa:2004cx}
\ba
\del_\mu j^\mu_{(b)}&=&\del_\mu j^\mu_{(l)}=\frac{n_f}{32\pi^2}\left[\frac{1}{2}\epsilon_{\mu\nu\rho\sigma}W^a_{\mu\nu}W^a_{\rho\sigma}\right],\\
   &=&\del_\mu K^\mu.
\ea
where
\ba
K^\mu&=&\frac{n_f}{16\pi^2}\epsilon_{\mu\nu\rho\sigma}\left[W^a_{\nu\rho}W^a_\sigma-\frac{2}{3}\epsilon_{abc}W^a_\nu W^b_\rho W^c_\sigma\right].
\ea
and $n_f$ is the number of fermion families, which we have equal to one for our simulation. We then find that the baryon number, $N_f=\int d^3x j^0_{(b)}$, is related to the Chern-Simons number, $N_{CS}=\int d^3x K^0$, as follows
\ba
N_f&=&N_{CS}.
\ea
The main aim of this paper is to confirm that this truly quantum mechanical relation holds in our numerical simulations, and to apply the method to a fast electroweak transition. We will also calculate the average Higgs field
\ba
\langle\phi^2\rangle=\frac{1}{\textrm{Vol}}\int d^3x\, \phi^\dagger\phi,
\ea
and the Higgs winding number,
\ba
N_W=\frac{1}{24\,\pi^2}\int d^3x\, \epsilon_{ijk}V^\dagger\partial_i VV^\dagger\partial_j VV^\dagger\partial_kV,\qquad V=\frac{1}{|\phi|}\phi.
\ea

\subsubsection*{For convenience: Partial charge conjugation transformation}
\label{app:nambyPamby}

On a lattice it is is more convenient to redefine the Fermi fields so that their kinetic terms are just the standard Dirac form, i.e. the fermion-gauge interactions are vector-like rather than chiral. This may be achieved by the following definitions \cite{Lee:1988ut,Anselm:1993uj,Aarts:1998td,Aarts:1999zn},
\ba
\Psi_R&=&\epsilon\mathcal{C}^{-1}\bar l_L^T\qquad\Rightarrow \bar\Psi_R=-l_L^T\mathcal{C}\epsilon^{-1},
          \quad l_L=-\epsilon\mathcal{C}^{-1}\bar\Psi_R^T,
          \quad\bar\l_L=\Psi^T_R\mathcal{C}\epsilon^{-1},\\
\Psi_L&=&q_L,\\
\chi_R&=&u_R,\\
\chi_L&=&\mathcal{C}^{-1}\bar e_R^T\qquad\Rightarrow e_R=\mathcal{C}^{-1}\bar \chi_L^T,\quad\bar e_R=-\chi_L^T\mathcal{C},\\
\xi_R&=&d_R,\\
\xi_L&=&\mathcal{C}^{-1}\bar \nu_R^T\qquad\Rightarrow \nu_R=\mathcal{C}^{-1}\bar\xi_L^T,\quad\bar \nu_R=-\xi_L^T\mathcal{C},
\ea
We then find that
\ba
\bar l_L\gamma^\mu\del_\mu l_L&\equiv&\bar\Psi_R\gamma^\mu\del_\mu\Psi_R,\\
\bar e_R\gamma^\mu\del_\mu e_R&\equiv&\bar\chi_L\gamma^\mu\del_\mu\chi_L,\\
\bar\nu_R\gamma^\mu\del_\mu\nu_R&\equiv&\bar\xi_L\gamma^\mu\del_\mu\xi_L,\\
\bar l_L\gamma^\mu\sigma^a W^a_\mu l_L&=&\bar\Psi_R\gamma^\mu\sigma^a W^a_\mu\Psi_R,
\ea
where some integration by parts has been done on the first three equations, leaving us with
\ba
S_F&=&-\int\; d^4x\;\left[ \bar\Psi\gamma^\mu D_\mu \Psi
                          +\bar\chi\gamma^\mu \del_\mu \chi+\bar\xi\gamma^\mu \del_\mu \xi\right],\\
S_Y&=&-\int\; d^4x\;\left[G^d\bar\Psi\phi P_R\xi+G^e\bar\chi\tilde\phi^\dagger P_R\Psi+G^u\bar\Psi\tilde\phi P_R\chi-\hat G^\nu\bar\xi\phi^\dagger P_R\Psi+h.c.\right].
\ea
The end result is that whereas we before had two left-handed doublets and four right-handed singlets, these are now collected into one full Dirac doublet and two Dirac singlets. The singlets only interact via the Yukawa term. It is straightforward to check that in this formulation one has
\ba\nonumber
\left(j^\mu_{(5)}\right)_{\rm C-conjugated}=\left(j^\mu_{(b)}+j^\mu_{(l)}\right)_{\rm Original}&=&i\left[-\bar\Psi\gamma^\mu\gamma^5\Psi+\bar\chi\gamma^\mu\gamma^5\chi+\bar\xi\gamma^\mu\gamma^5\xi\right].\\
\ea

\section{Modelling the bosons and fermions}
\label{sec:modelling}

The continuum model above is discretized on the lattice as described in App.~\ref{app:latticeAction}. For the bosonic fields the classical Hamiltonian equations are derived, with the understanding that fermion bilinears are to be represented by quantum averages. The fermion evolution equations are also found in the usual way by variation of the action, and amount to a linear Dirac equation in a time- and space-dependent bosonic background. 

One issue to address are the lattice fermion doublers. For every physical fermion mode, the standard lattice discretization prescription generates a set of fifteen doublers, which count as real degrees of freedom at finite lattice spacing \cite{Smit:2002ug}; in the context of the anomaly these are particularly troublesome. Since they contribute with opposite signs, the doublers cancel the anomaly from the physical fermions, and  set the total anomaly to zero \cite{Bock:1994fu} (see also Fig.~\ref{fig:doublers}).

Fortunately, a simple approach exists to breaking the symmetry between these doublers, namely adding a Wilson term to the lattice action (\ref{eq:latticeaction}). As was the case in 1+1 dimensions \cite{Aarts:1998td,Aarts:1999zn,Saffin:2011kc}, we find that it is sufficient to add a spatial Wilson term to cancel the space-like doublers, and to not initialize the time-like doublers. For small enough timestep in the numerical evolution, the time-like doublers stay unexcited, at least for the duration of our simulations (see also section \ref{sec:sphalerons}). 

The next step is to replace quantum averages by ensemble averages, as discussed in detail in \cite{Borsanyi:2008eu,Saffin:2011kc}. The process starts by noting that the continuum canonical anti-commutation relations for continuum fermions are
\ba
\{\psi_\alpha(t,\underline x),\psi^\dagger_\beta(t,\underline x')\}=\delta_{\alpha\beta}\delta(\underline x-\underline x'),
\ea
and we may expand the field operator as
\ba\nonumber
\psi(t,\underline x)&=&\sum_s\int\;\frac{d^3p}{(2\pi)^3}\frac{1}{2\omega_p}\left[b_s(\underline p)U_s(\underline p)e^{ip.x}+d^\dagger_s(\underline p)V_s(\underline p)e^{-ip.x}\right],
\ea
in terms of the creation and annihilation operators. The anti-commutation relation are then equivalent to imposing
\ba
\{b_r^\dagger(\underline p),b_s(\underline p')\}&=&(2\pi)^3(2\omega_p)\delta_{rs}\delta(\underline p-\underline p'),\\
\{d_r^\dagger(\underline p),d_s(\underline p')\}&=&(2\pi)^3(2\omega_p)\delta_{rs}\delta(\underline p-\underline p'),
\ea
and we may calculate bi-linears such as
\ba
\langle\;|\bar\psi(x)\psi(y)|\;\rangle&=&\sum_s\int\;\frac{d^3p}{(2\pi)^3}\frac{1}{2\omega_p}
              \bar V_s(\underline k)V_s(\underline k)e^{ip.(x-y)}\\
\label{eq:biLinearQuant}
  &=&-4\mu\int\;\frac{d^3p}{(2\pi)^3}\frac{1}{2\omega_p}e^{ip.(x-y)},
\ea
where $\mu$ is the fermion mass; it is these kinds of quantities that we aim to reproduce using ensemble averages.

To model the quantum averages we introduce two ensembles of fermions, M(ale) and F(emale), according to the mode expansion
\ba
\label{eq:ferminit}
\psi_{M,F}(t,\underline x)&=&\frac{1}{\sqrt{2}}\sum_s\int\;\frac{d^3p}{(2\pi)^3}\frac{1}{2\omega_p}\left[\xi_s(\underline p)U_s(\underline p)e^{ip.x}
                \pm\eta_s(\underline p)V_s(\underline p)e^{-ip.x}\right],
\ea
and where the exact same random numbers $\xi$, $\zeta$ are used in a given Male and Female pair.
By then requiring that the variables $\xi$ and $\eta$ satisfy the ensemble average relations
\ba
\langle\xi_r(\underline p)\xi^\star_s(\underline p')\rangle_e&=&(2\pi)^3(2\omega_p)\delta_{rs}\delta(\underline p-\underline p'),\\
\langle\eta_r(\underline p)\eta^\star_s(\underline p')\rangle_e&=&(2\pi)^3(2\omega_p)\delta_{rs}\delta(\underline p-\underline p'),
\ea
(where $\delta(\underline p-\underline p')$ is understood to refer to the lattice version of the Dirac delta), we may for example calculate
\ba
\langle\bar\psi_M(x)\psi_F(y)\rangle
     &=&\frac{1}{2}\sum_s\int\frac{d^3p}{(2\pi)^3}\frac{1}{2\omega_p}\left[
        \bar U_s(\underline p) U_s(\underline p)e^{-ip.(x-y)}
       -\bar V_s(\underline p) V_s(\underline p)e^{ip.(x-y)}\right]\nonumber\\
   &=&+4\mu\int\frac{d^3p}{(2\pi)^3}\frac{1}{2\omega_p}e^{ip.(x-y)},
\ea
which leads to the simple prescription
\ba
\langle\;|\bar\psi(x)\psi(y)|\;\rangle \rightarrow -\frac{\langle\bar\psi_M(x)\psi_F(y)+(M\leftrightarrow F)\rangle_e}{2}.
\ea
For more details of the method see \cite{Borsanyi:2008eu}.

The mode expansion described above constitutes the initial conditions for our fermions, which then amounts to generating sets of $\eta_k,\xi_k$ and inserting them into eq. (\ref{eq:ferminit}). In the case of a general Yukawa interaction and CKM mixing matrix, one has to take care to diagonalize the fermions first to initialize the mass eigenmodes; this is explained in more detail in App. \ref{app:ferminit}. For the simulations presented here, we will restrict ourselves to Yukawa couplings proportional to the identity (see also App. \ref{app:latticeAction}) 
\ba
G^e=G^u=G^d=-G^\nu=\lambda_{\rm yuk}.
\ea

\section{Sphaleron transitions}
\label{sec:sphalerons}

\begin{figure}
  \centering
  \includegraphics[width=0.65\textwidth]{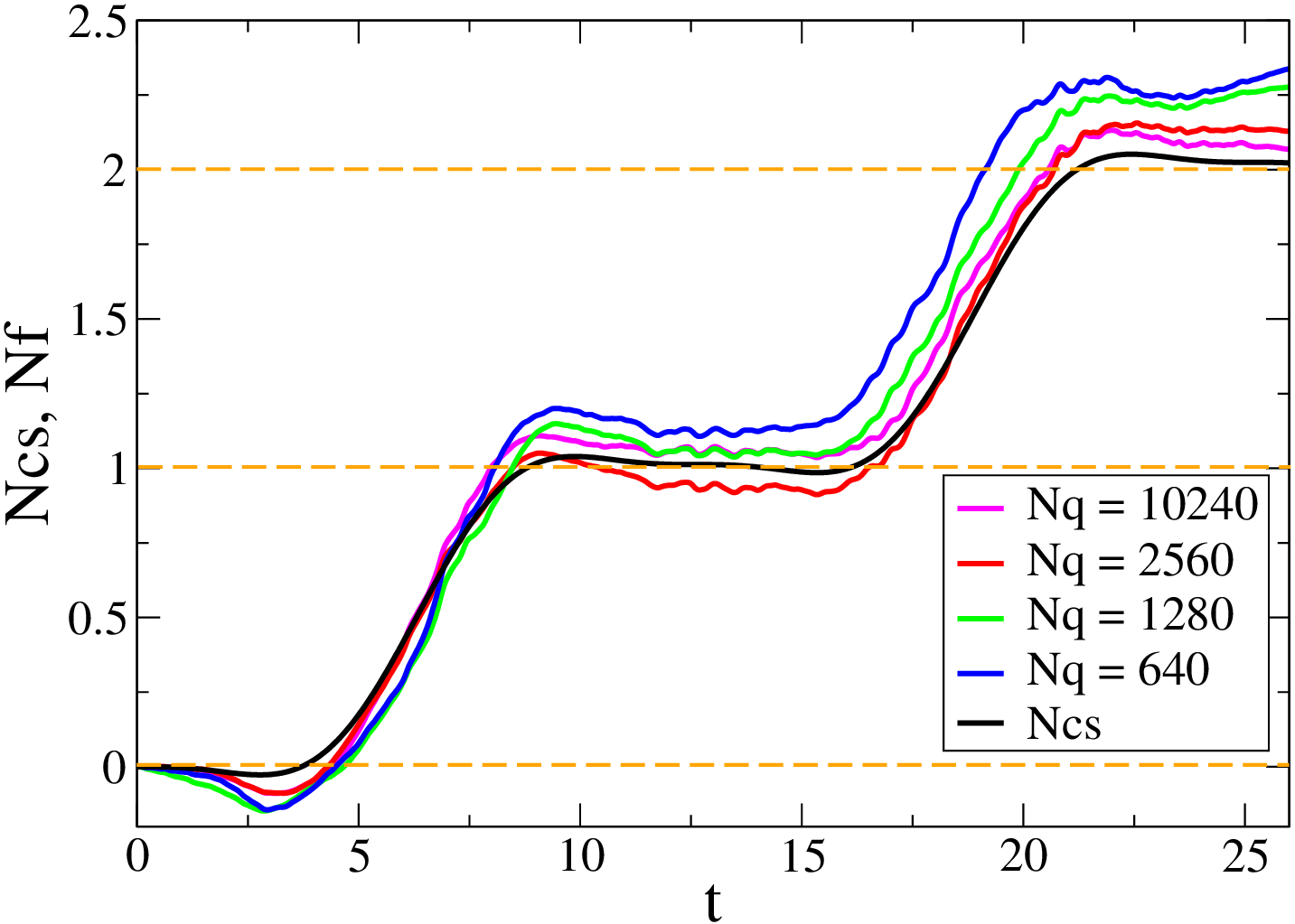}
  \caption{\label{fig:fig1} The fermion number compared to the Chern-Simons number in a handmade sphaleron transition, for different sizes of the fermion ensemble, $N_q$. There is clear convergence to the exact result, which is within statistical error bars (not shown) for all values of the ensemble here. $am_H=0.42$. }
\end{figure}

Now that we have the full prescription for evolving the system of bosons and fermions, we wish to test whether the method is reliable for the anomaly dynamics driven by sphaleron processes. We are mainly interested in whether the numerical fermions react in the correct way as the SU(2) gauge field undergoes a change in winding. To that end we shall start by evolving the Higgs and gauge fields by hand, rather than through their equations of motion, so we need to know how to construct a winding event. This is the method used in 1+1 dimensions \cite{Aarts:1998td,Aarts:1999zn,Saffin:2011kc}, and here we extend the construction to the SU(2) sphaleron following the techniques of \cite{Manton:1983nd} (see also \cite{Klinkhamer:2003hz} for a nice exposition). The detailed implementation can be found in App. \ref{app:HMsphalerons}.

The end result is a family of configurations starting and ending in a pure-gauge vacuum of the gauge-Higgs theory, which takes the fields through a sphaleron transition. In this way there is a relative winding number between the vacua, both in terms of winding in the Higgs field, and the Chern-Simons number of the gauge field. On this bosonic background we evolve the fermions according to their dynamical evolution equations, and test that they respond in the expected way and acquire a non-zero fermion number in accordance with the anomaly equation. The fermions are initialized in their vacuum as described in App. \ref{app:ferminit}.

\begin{figure}
  \centering
\includegraphics[width=0.75\textwidth]{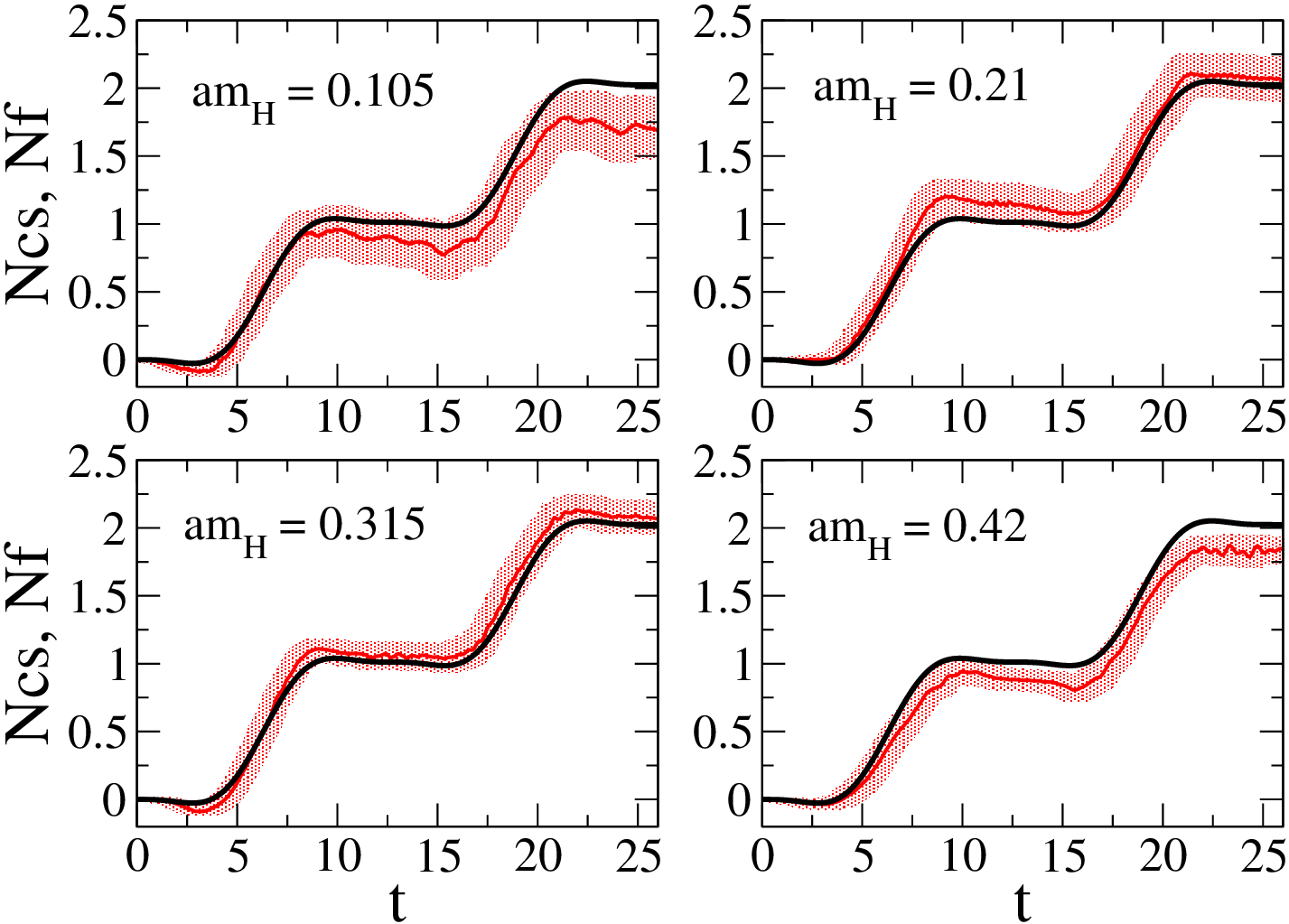}
  \caption{\label{fig:fig2} Handmade sphaleron transitions for different values of the lattice spacing $am_H=0.105-0.42$. The lattice spacing dependence is well under control. The lattice size was kept constant at $n_x=32$. The band is the statistical errors.}
\end{figure}

For the moment, we will set the Yukawa couplings to zero. We generate a double-sphaleron trajectory on a $n_x^3=32^3$ lattice with $m_H/m_W=2$, a timestep $a_t/a_x=0.05$ a lattice spacing $am_H=0.42$ and a Wilson coefficient $r=1.0$. Fig. \ref{fig:fig1} shows Chern-Simons number $N_{\rm cs}$ in time (black) and the combination $(B+L)/2$ increasing the fermion ensemble. The coloured lines show the ensemble averaged anomaly for $N_q=640,1280,2560$ and our largest ensemble $N_q=10240$. We see that the anomaly is very well reproduced at the ten percent level for these ensemble sizes, and that this can be improved upon in a straightforward way. For a $32^3$ lattice, numerical efficiency is only gained over the ``non-statistical'' approach for ensemble sizes less that $32^3=32768$. 

\begin{figure}
  \centering
  \includegraphics[width=0.65\textwidth]{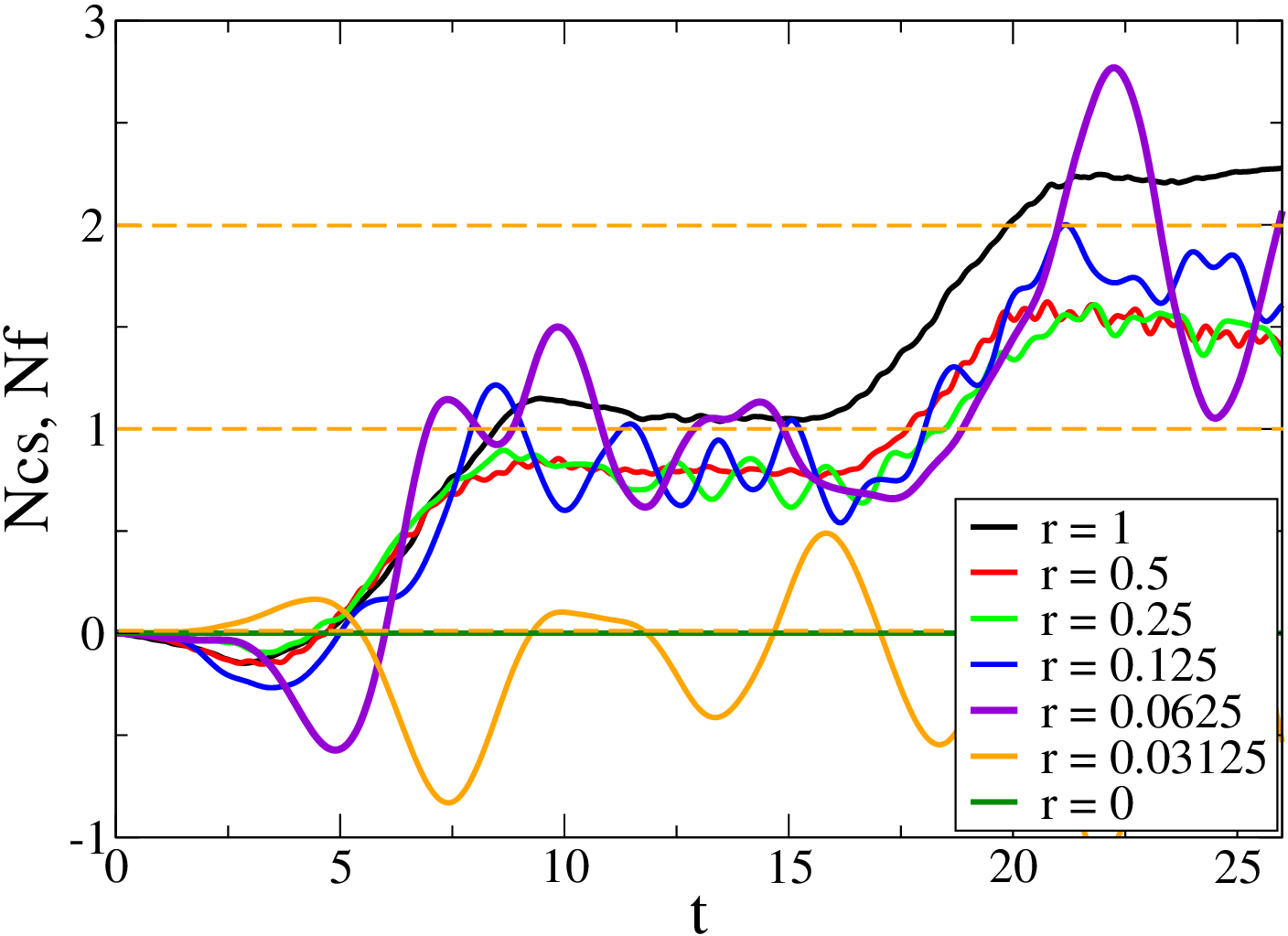}
  \caption{\label{fig:doublers} The anomaly as the spatial Wilson coefficient is varied. }
\end{figure}

The next test is to vary the lattice spacing (for fixed number of lattice points), shown in Fig.~\ref{fig:fig2}. We see that the anomaly is well reproduced for all lattice spacings, although the $am_H=0.105$ is a little low, possibly because the physical volume is quite small $n_xam_H=3.36$. Also the largest lattice spacing $am_H=0.42$ is about 5 percent low, which may in turn be put down to discretization errors. 
We note that the anomaly is expected to be one of the ``hardest'' observables to get right, as it does not involve any volume averaging (such as for instance particle number in a homogeneous background would \cite{bergesgelfand}). Also, as can be deduced from the level-crossing picture of the anomaly, all momentum modes contribute to the anomaly, both in the IR and the UV. Hence the statistical averaging has to be accurate for all modes independently, to get the correct anomaly out. It is not possible to leave out some of the UV modes, as if one were only interested in the IR physics. In this light, we consider it remarkable that our simulations reproduce the anomaly so well.

In Fig.~\ref{fig:doublers} we show how the anomaly depends on the coefficient of the Wilson term. As mentioned, for $r=0$ lattice doublers are expected to pair up and cancel out the anomaly, whereas for $r=1$, the anomaly is manifest. We indeed see that the anomaly is identically zero for $r=0$, but also that once $r>0.25$ the anomaly is largely unaffected by $r$ (within statistical errors). This is as we would have expected. Also note that the time-like doublers are not cancelled by a Wilson term, but that these are not excited enough to cause problems.

\begin{figure}
  \centering
  \includegraphics[width=0.75\textwidth]{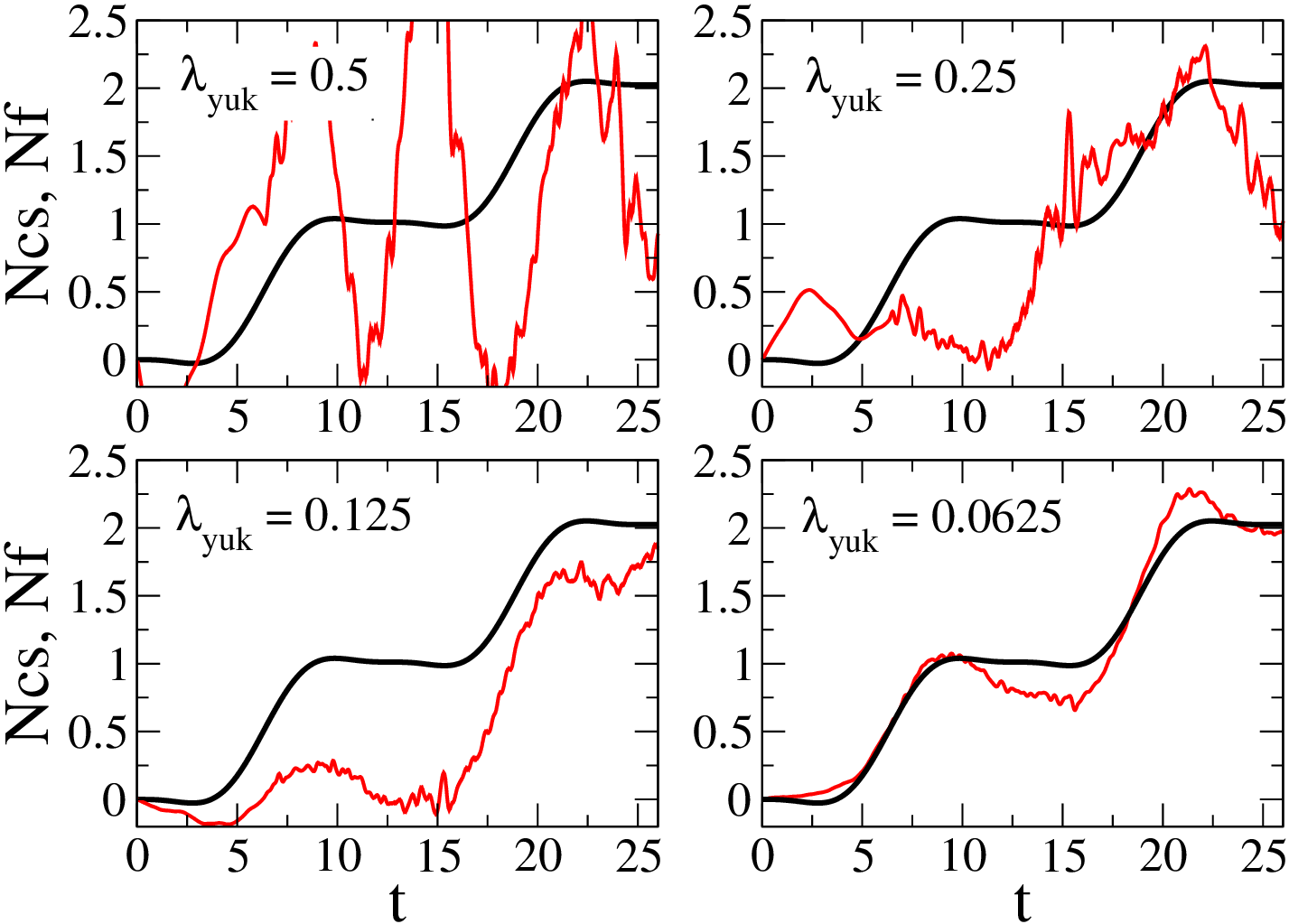}
  \caption{\label{fig:yukspha32} Handmade sphaleron simulations, varying the Yukawa coupling. $N_q=10240$. $n_x=32$, $am_H=0.42$. The Yukawa couplings correspond to fermion masses of $m_f=87,44,22,11\,$GeV, respectively.}
\end{figure}

\begin{figure}
  \centering
 \includegraphics[width=0.75\textwidth]{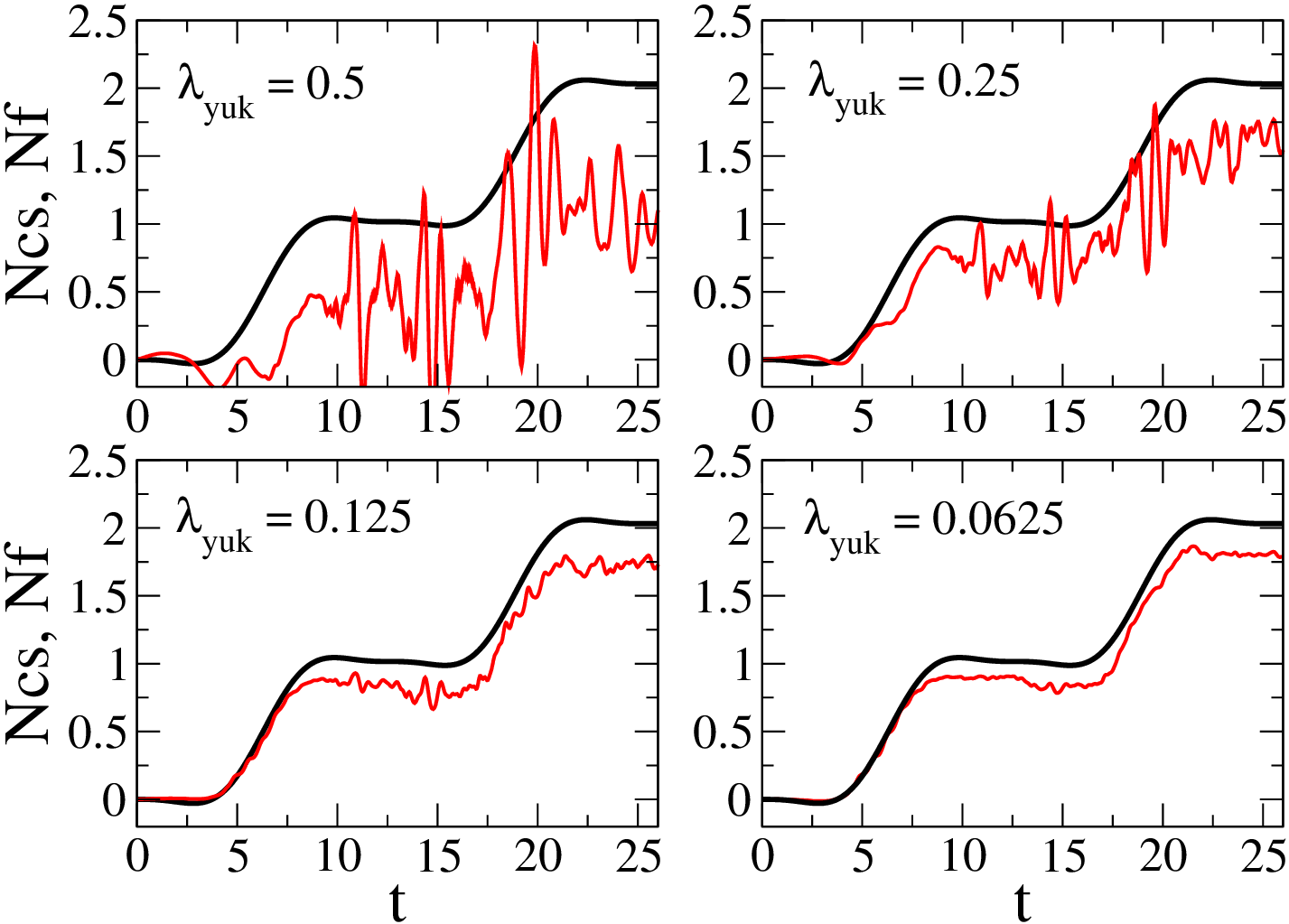}
  \caption{\label{fig:yukspha16} Handmade sphaleron simulations for a smaller lattice $16^3$ with much larger statistics $N_q=81920$.}
\end{figure}

We then consider the case where scalars are directly coupled to the fermions, i.e. the fermions are massive. As shown in Fig. \ref{fig:yukspha32}, at finite Yukawa coupling convergence slows down with increasing coupling. This is completely analogous to the situation in 1+1 dimensions \cite{Saffin:2011kc}, and the solution is also the same: increase the size of the ensemble. $N_q=10240$ is again sufficient for $\lambda_{\rm yuk}<0.1$, corresponding to a fermion mass of $\lambda_{\rm yuk}v/\sqrt{2}\simeq 17\,$GeV, i.e. all Standard Model quarks except the top. But to include the top, $\lambda_{\rm yuk}\simeq 1$, we need many more realizations. We did a test at the smaller lattice size of $n_x^3=16^3$ with $N_q=8\times 10240$ and indeed found that the anomaly is still present, shown in Fig.~\ref{fig:yukspha16}. Such simulations suffer from having small physical volume, and the fermion number is a little low. We also explicitly checked that decreasing the timestep and/or increasing the Wilson coefficient $r$ did not improve convergence.
\cite{Manton:1983nd}
We can conclude from these tests that the fermion ensemble method works even for very subtle quantum observables like the anomaly, and that we need $N_q=\mathcal{O}(10^4)$ to get a measurement at the ten percent accuracy level, at zero and small Yukawa couplings, up to a fermion mass of a few GeV. A larger ensemble is required to consistently include heavy quarks, i.e. the top. Lattice issues like lattice spacing dependence and fermion doublers are well under control. We will now proceed to also make the bosonic fields dynamical and include the fermion back-reaction.

\section{Including fermion backreaction: tachyonic electroweak transition}
\label{sec:tacres}

In the tests with handmade sphalerons presented above, the fermions were just ``along for the ride'', to see how well the fermion equations of motion behave in a time-dependent bosonic background. But for baryogenesis, we need the bosonic fields to be dynamical, too, and for the fermions to back-react. Indeed, in the SM CP-violation {\it is } a fermion back-reaction effect.

As a testing ground for this, we consider a cold tachyonic electroweak transition, where the Higgs field experiences a fast quench. The Higgs potential flips instantaneously as
\ba
V(\phi) = +\mu^2\phi^\dagger\phi,\quad (t<0)\quad \rightarrow\quad \lambda\left(\phi^\dagger\phi-\frac{v^2}{2}\right)^2, \quad(t>0).
\ea
The Higgs field is generated by Monte-Carlo sampling of a classical ensemble, reproducing the quantum correlation functions at zero temperature in the symmetric phase \cite{GarciaBellido:2002aj,Smit:2002yg,Tranberg:2003gi}. The fermion fields are generated as an ensemble of $N_q$ vacuum realizations at $\phi=0$, as above. The initial gauge fields are then found by setting $A_\mu(t=0)=0$, and solving the lattice Gauss law in the background of the generated Higgs and averaged fermion fields.

The dynamics start at the moment when the Higgs potential is flipped, and the low-momentum modes $k<\mu$ experience a tachyonic (or spinodal) instability, and start growing exponentially. This leads to a strongly out-of equilibrium electroweak transition, where large particle numbers, effective diffusion of Chern-Simons number and in the presence of CP-violation, baryogenesis can occur \cite{GarciaBellido:1999sv,GarciaBellido:2003wd,Tranberg:2003gi,Tranberg:2009de}. In the present work, we have only implemented one generation of Standard Model fermions, and so we are not able to include CP-violation via the CKM matrix. We will leave that for future work.

As in \cite{Tranberg:2003gi} we will monitor the volume-averaged Higgs field, the Chern-Simons number and the Higgs winding number. But rather than infer the baryon and lepton number through the anomaly equation, we now calculate it directly. We note that the anomaly equation states that changes in the Chern-Simons and Baryon/Lepton number obey
\ba
\Delta B = \Delta L=\Delta N_{\rm cs}\quad \rightarrow \quad \Delta N_{\rm cs}=\frac{\Delta B+\Delta L}{2}.
\ea

\begin{figure}
  \centering
  \includegraphics[width=0.65\textwidth]{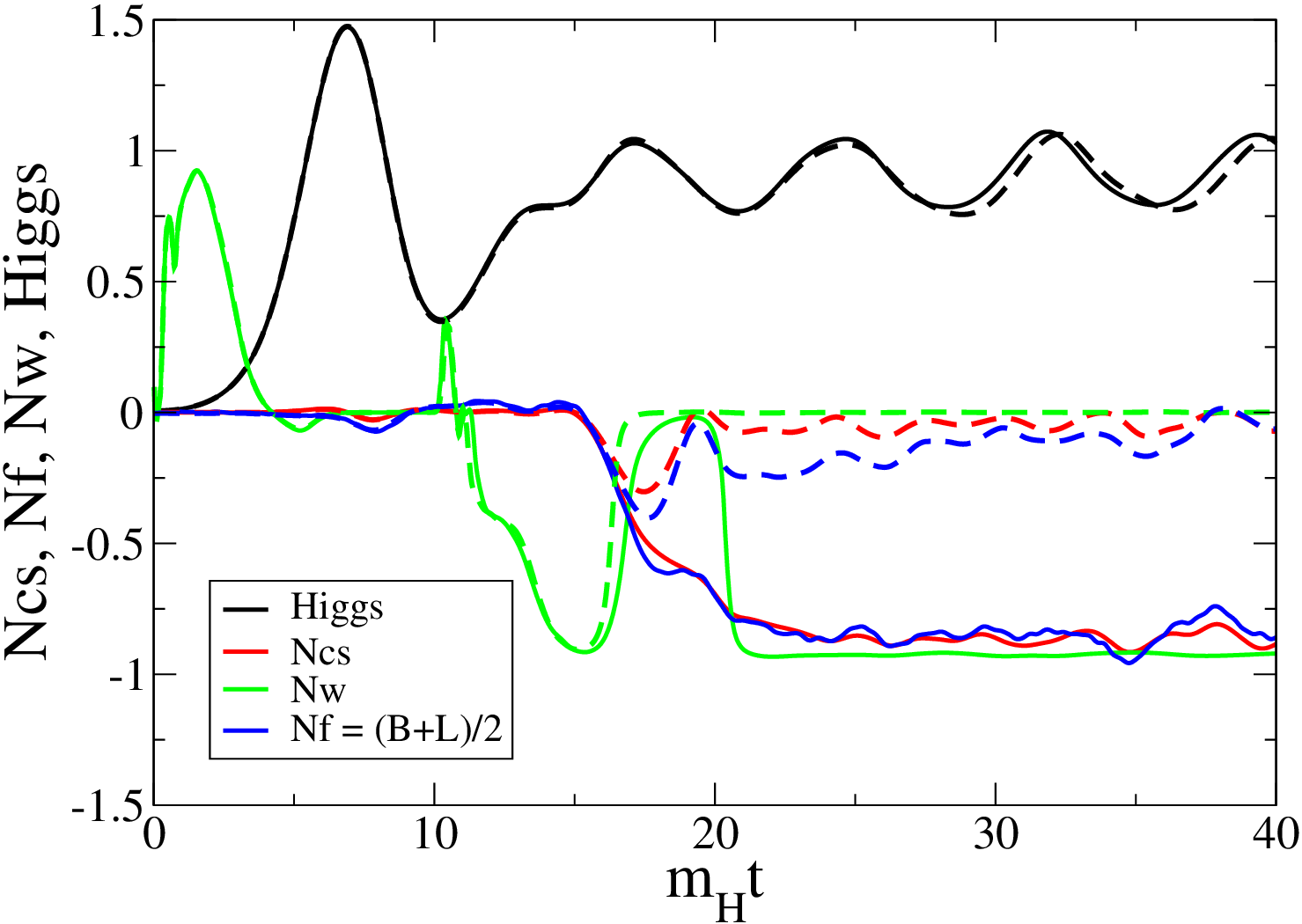}
  \caption{\label{fig:tacback} A tachyonic transition with (full) and without (dashed) fermion back-reaction, starting from the exact same initial condition. Shown are the trajectories of the Chern-Simons number (red), the average Higgs field squared (black) the Higgs winding number (green) and the fermion number (blue).\newline\newline}
\end{figure}

Fig.~\ref{fig:tacback} shows a tachyonic run where the fermions do not back-react (dashed) and one where they do (full). Yukawa couplings are zero, and $N_q=5120$.  We see that the Higgs field (black) ``falls off the hill'' and oscillates around its broken phase minimum. Meanwhile, the Chern-Simons number (red) and Higgs winding number (green) bounce around out of equilibrium, but eventually settle down in a common minimum. To check our numerics, we also monitored Gauss law, which throughout is conserved to computer accuracy. As for the sphalerons above, we see that the fermions (blue) reproduce the anomaly rather well, however, the gauge and Higgs fields are now dynamical.

What is particularly interesting is the effect of having fermions affecting the bosonic dynamics. At early times, the evolution of the bosonic observables is unchanged. But then around $m_Ht=15$, the trajectories of both fermion number and Chern-Simons number diverge, depending on whether the bosons feel the fermions, eventually ending up at different minima, and hence a different number of fermions.

\begin{figure}
  \centering
  \includegraphics[width=0.65\textwidth]{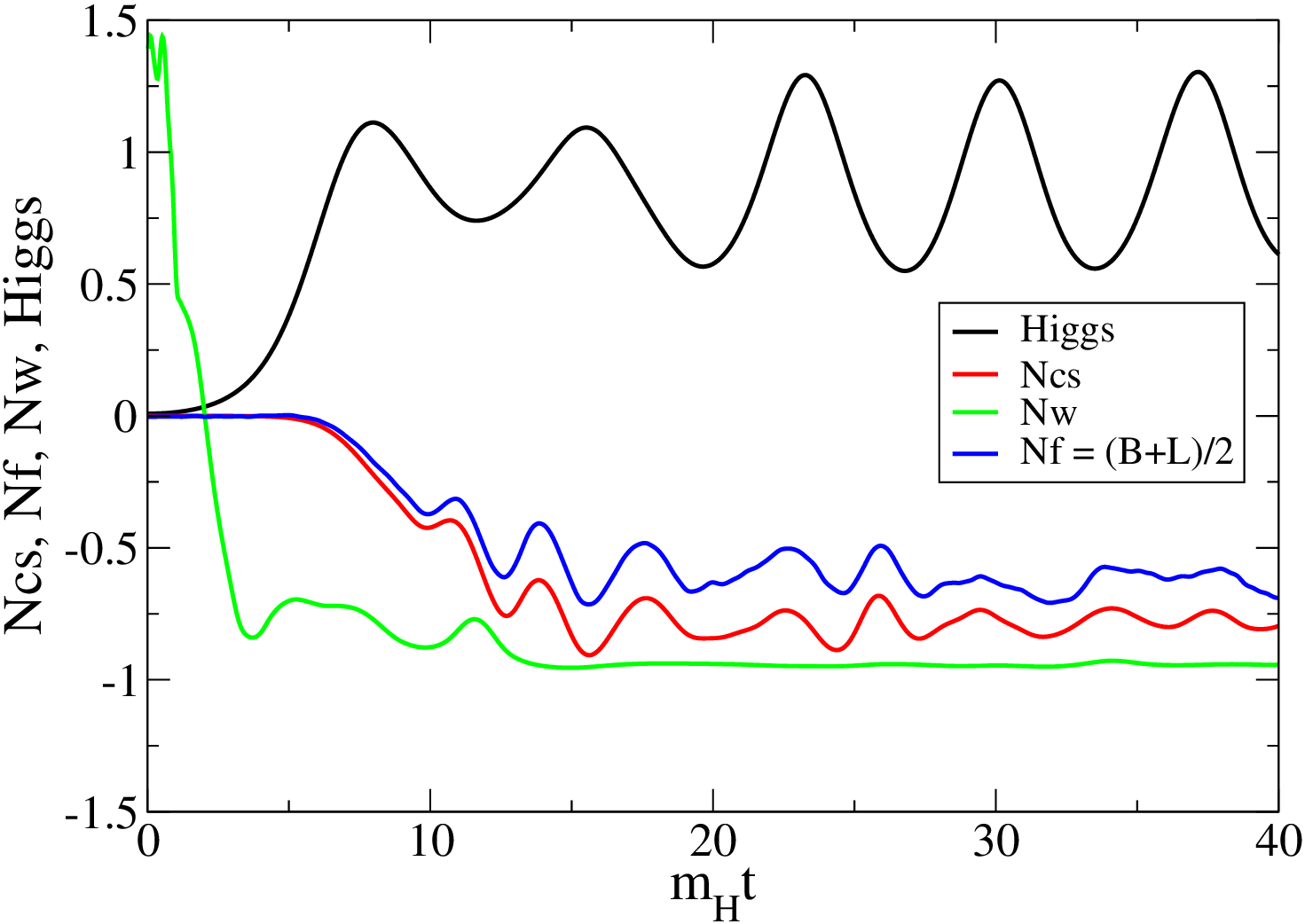}
  \caption{\label{fig:tacbackyuk} A full tachyonic run including fermion backreaction and Yukawa coupling. The fermion mass is $m_f\simeq 1.4\,$GeV.}
\end{figure}

Finally, in Fig.~\ref{fig:tacbackyuk} we show a run with small enough $\lambda_{\rm yuk}=1/128$, that $N_q=10240$ gives reliable convergence. This corresponds to a $1.4\,$GeV fermion, i.e. slightly heavier than the charm quark. As for the sphaleron runs, we see that the agreement is only approximate, but clearly the anomaly is reproduced, improveable with larger statistics.

\section{Conclusions}
\label{sec:conclusion}

While there have been many simulations of the bosonic fields in the Standard Model, this is the first that has included the fermionic sector, rather than integrating them out and/or treating them as passive observers to the gauge-scalar dynamics. This has been achieved by applying the technique pioneered by Bors\'{a}nyi and Hindmarsh \cite{Borsanyi:2008eu}, where the many mode-functions of the fermion fields, required for a full quantum treatment, are replaced by an average over a statistical ensemble of Male and Female fields.

We have performed a detailed study of the lattice parameters, and found that the fermion number does indeed reproduce the quantum anomaly, with the fermion number tracking the Chern-Simons number through sphaleron transitions. We also simulated a fully dynamical fast electroweak transition including fermions, and found that the back-reaction of fermions can indeed significantly alter the evolution of the gauge fields. 

From a technical point of view, the numerical effort is significant, in that implementing $N_q$ fermion realizations on a $n_x^3$ lattice requires $n_x^3\times N_q\times 1.6\,$kB, so that for our single generation, no-colour full run of Fig.~\ref{fig:tacback} we use approximately $270\,$GB. It is however important to note that the anomaly is probably one of the most sensitive observables, since it requires every single lattice mode to be very precisely reproduced. This means that for studying the (IR) dynamics of electroweak baryogenesis, we can probably make do with a somewhat smaller $N_q$. Alternatively, one may consider returning to the non-ensemble approach \cite{Aarts:1998td,Aarts:1999zn}, where one is however hampered by the need for large lattices to correctly simulating tachyonic transitions.

The next step is now to study dynamics both for baryogenesis and electroweak preheating; including all three generations, mixing and CP-violation for both the quark (CKM) and lepton (PMNS) sector as relevant for the most minimal version of Cold Electroweak Baryogenesis; but also for electroweak baryogenesis in more general setups like Standard Model + second Higgs doublet or Standard Model + scalar singlet. Demonstrating that this is a realistic aim (also in practice) is perhaps the most significant result of the present work.

\acknowledgments

P.S. would like to acknowledge STFC for financial support A.T. is supported by the Carlsberg Foundation. The numerical work was conducted at the Finnish supercomputing center CSC.

\appendix

\section{Conventions}
\label{app:conventions}

We consider 3+1D Minkowski space, with the metric $(-,+,+,+)$, and Dirac algebra
\ba
\{\gamma^\mu,\gamma^\nu\}=2\eta^{\mu\nu},\quad \bar\psi=i\psi^\dagger\gamma^0,\quad \gamma_5=-i\gamma^0\gamma^1\gamma^2\gamma^3,
\ea
with in the Weyl representation
\ba
\gamma^0=
\left(\begin{array}{cc}0  & i\Id\\i\Id & 0\end{array}\right)
,\quad
\gamma^j=
\left(\begin{array}{cc}0 & i\sigma^j\\-i\sigma^j & 0\end{array}\right)
,
\gamma_5=
\left(\begin{array}{cc}\Id & 0\\0 & -\Id\end{array}\right),\\
\mathcal{C}=
\left(\begin{array}{cc} -i\sigma^2 & 0\\0 & i\sigma^2 \end{array}\right)=-i\gamma^0\gamma^2
,\quad CC^\dagger=\Id,\quad C^\dagger=-C,\\
\gamma^{\mu T}=-C\gamma^\mu C^{-1},
\quad \gamma^{\mu \dagger}=\gamma^0\gamma^\mu \gamma^0=\gamma_\mu,\\
\epsilon=i\sigma^2, \quad\epsilon^{-1}\sigma^a\epsilon=-\sigma^{a\star},
\ea
\ba
\psi_R=\frac{1}{2}(1+\gamma^5)\psi, \quad\psi_L=\frac{1}{2}(1-\gamma^5)\psi.
\ea
The charge-conjugate scalar doublet is given by the complex conjugate
\ba
\phi^c&=&\epsilon\phi^\star=\tilde{\phi}.
\ea

\section{Lattice action}
\label{app:latticeAction}

The theory is discretized on a lattice with spatial lattice spacing $a_x$, timelike spacing $a_t=dta_x$. We find it convenient to work with the following rescaled and re-defined variables
\ba
W_{\rm lattice}&\equiv&  gW_{\rm continuum},\\
\phi(x)&=&\left(\begin{array}{c}\phi_1(x)\\\phi_2(x)\end{array}\right)\rightarrow
\Phi(x)=\sqrt{\lambda}a\left(\begin{array}{cc}\phi_2^*(x)&\phi_1(x)\\-\phi_1^*(x)&\phi_2(x)\end{array}\right)=\sqrt{\lambda}a\left(\tilde{\phi}(x),\phi(x)\right),\\
\Psi(x)&\rightarrow& a^{3/2}\Psi(x),\quad \chi(x)\rightarrow a^{3/2}\chi(x),\quad \xi(x)\rightarrow a^{3/2}\xi(x),
\ea
thus making them dimensionless. Then we have the link variables,
\ba
U_0(x)&=&\exp(-ia_t \sigma^aW^a_0(x)/2),\\
U_i(x)&=&\exp(-ia_\mu \sigma^aW^a_i(x)/2),\\
U_{ij}(x)&=&U_i(x)U_j(x+i)U^\dagger_i(x+j)U^\dagger_j(x)=\exp\left(-i\frac{1}{2}a^2\sigma^aW^a_{ij}+...\right)
\ea
and corresponding electric fields
\ba
E_i(x)&=&-U_i(x)U_i^\dagger(x+0),\qquad E_i^a(x)=\Tr\, i\sigma^a E_i(x),\\
\tilde{E}_i(x)&=&-U_i^\dagger(x+0)U_i(x), \qquad \tilde{E}_i^a(x)=\Tr\, i\sigma^a \tilde{E}_i(x).
\ea
We need the derivatives, where we will not include the lattice spacings (they are compiled into lattice coefficients, see below)
\ba
\label{eq:latticeDerivs}
\pi(x)&=&\Phi(x+0)-\Phi(x),\\\nonumber
D_\mu\Phi&=&U_\mu(x)\Phi(x+\mu)-\Phi(x),\\\nonumber
D'_\mu\Phi&=&\Phi(x)-U^\dagger_\mu(x-\mu)\Phi(x-\mu),\\\nonumber
D'_\mu D_\mu\Phi&=&U_\mu(x)\Phi(x+\mu)-2\Phi(x)+U^\dagger(x-\mu)\Phi(x-\mu),\\
\tilde D_\mu&=&\frac{1}{2}\left[D_\mu+D'_\mu\right].
\ea
Note that we are using the compact formulation of the gauge action, and have chosen to evolve using temporal gauge $W^a_0=0$ ($U_0=\Id$). In order to remove the spatial fermion doubler modes we will employ Wilson fermions in space, for which we need the Wilson term \cite{Aarts:1998td,Aarts:1999zn,Wilson:1975id}
\ba
\bar{\Psi}W\Psi=-\half \frac{r}{a^4}\, \bar{\Psi}D'_i D_i\Psi,\quad \bar{\chi}W\chi=-\half \frac{r}{a^4}\, \bar{\chi}\partial'_i \partial_i\chi,\quad \bar{\xi}W\xi=-\half \frac{r}{a^4}\, \bar{\xi}\partial'_i \partial_i\xi,
\ea
where $r$ is a constant parameter. We will also introduce the quantities
\ba
\beta_G^t = \frac{4}{g^2}\frac{a}{a_t},\quad \beta_G^s = \frac{4}{g^2}\frac{a_t}{a},\quad \beta_H^t = \frac{1}{\lambda}\frac{a}{a_t}, \quad \beta_H^s = \beta_R = \frac{1}{\lambda}\frac{a_t}{a},\quad
\frac{(am_H)^2}{4}=\frac{\lambda (av)^2}{2}.
\ea
The action is then given by
\ba
S=S_H+S_A+S_F+S_W+S_Y,
\label{eq:latticeaction}
\ea
with 
\ba
S_H &=& \sum_{x,t} \frac{\beta_H^t}{2}\Tr\, \left[(D_0\Phi)^\dagger D_0\Phi\right]-\frac{\beta_H^s}{2}\Tr\left[(D_i\Phi)^\dagger D_i\Phi\right]-\beta_R\left(\frac{1}{2}\Tr\, \Phi^\dagger\Phi-\frac{(am_H)^2}{4}\right)^2,\nonumber\\
\\
S_A &=& \sum_{x,t}\beta_G^t\sum_i\left(1-\frac{1}{2}\Tr \, U_{0i,x}\right)-\frac{\beta_G^s}{2}\sum_{ij}\left(1-\frac{1}{2}\Tr\,  U_{ij,x}\right),\\
S_F&=&\sum_{x,t}-\left[\bar{\Psi}\gamma^{0}\tilde{D}_{0}\Psi+\bar{\chi}\gamma^{0}\tilde{\partial}_{0}\chi+\bar{\xi}\gamma^{0}\tilde{\partial}_{0}\xi\right]-\frac{a_t}{a}\left[\bar{\Psi}\gamma^{i}\tilde{D}_{i}\Psi+\bar{\chi}\gamma^{i}\tilde{\partial}_{i}\chi+\bar{\xi}\gamma^{i}\tilde{\partial}_{i}\xi\right],\\
S_W&=&\sum_{x,t}\frac{ra_t}{2a}\, \left[\bar{\Psi}D_i^{'}D_i\Psi+\bar{\chi}\partial_i^{'}\partial_i\chi+\bar{\xi}\partial_i^{'}\partial_i\xi\right],
\ea
To which we add the Yukawa terms, now written in terms of the matrix $\Phi$, and where we for simplicity impose that
\ba
G^u=G^{e*},\qquad G^d=-G^{\nu*},
\ea
and in addition set
\ba
G^u=G^d=\lambda_{yuk}.
\ea
Then we have, using
\ba
\beta_{Y}=\frac{a_t}{a}\frac{1}{\sqrt{\lambda}}.
\ea
the addition to the action
\ba
S\rightarrow S+S_{yuk},\qquad S_{yuk}=-\sum_{x,t}
\left(\bar{\Psi}_a\Phi_{ab}(\chi,\xi)_b^T+(\bar{\chi},\bar{\xi})_b\Phi^\dagger_{ba}\Psi\right).
\ea
The equations of motion follow by variation. For the Higgs field
\ba
\partial_t\partial_t'\Phi= \frac{\beta_H^t}{\beta_{H}^s}D_i'D^i\Phi-2\frac{\beta_R}{\beta_H^t}\left(\frac{1}{2}\textrm{Tr}\Phi^\dagger\Phi-\frac{v^2}{2}\right)\Phi-2\frac{\beta_Y}{\beta_H^t}\delta\Phi=0,
\ea
where
\ba
\delta\Phi = \delta\phi_0 I+\delta\phi_a i \sigma^a,
\ea
and
\ba
\delta\phi_0+i\delta\phi_3&=&-\langle\bar{\chi}^M\Psi_u^F+\bar{\Psi}_d^M\xi^F+(M\leftrightarrow F)\rangle,\\
\delta\phi_2+i\delta\phi_1&=&-\langle\bar{\xi}^M\Psi_u^F-\bar{\Psi}_d^M\chi^F+(M\leftrightarrow F)\rangle,
\ea
are ensemble average over fermion realizations. 
For the gauge field we have
\ba
\partial_t'E_n^a(y)&=&\frac{\beta_g^s}{\beta_g^t}\sum_m D_m^{ab'}\textrm{Tr}\left[i\sigma^bU_{y,m}U_{y+m,n}U_{y+n,m}^\dagger U_{y,n}^\dagger\right]\nonumber\\
&&-\frac{2\beta_H^s}{\beta_{g}^t}\textrm{Tr}\left[i\sigma^a\Phi_y(U_{y,n}\Phi_{y+n})^\dagger\right]
-\frac{a_t}{a\beta_{g}^t}\left[\bar{\Psi}_y\gamma^n i \sigma^a U_{y,n}\Psi_{y+n}+\bar{\Psi}_{y+n}\gamma^nU^\dagger_{y,n}i\sigma^a\Psi_y\right]\nonumber\\
&&+\frac{ra_t}{a\beta_g^t}\left[\bar{\Psi}_yi \sigma^a U_{y,n}\Psi_{y+n}-\bar{\Psi}_{y+n}U^\dagger_{y,n}i\sigma^a\Psi_y\right],
\ea
and we have Gauss law
\ba
D_n^{ab'}E_n^b(y)=\frac{2\beta_H^t}{\beta_G^t}\textrm{Tr}i\sigma^a\Phi_y\Phi^\dagger_{y+0}-\frac{1}{\beta_G^t}\left[\bar{\Psi}_y\gamma^0 i\sigma^a\Psi_{y+0}+\bar{\Psi}_{y+0}\gamma^0i\sigma^a\Psi_y\right].
\ea
Finally, the fermion fields evolve according to
\ba
\frac{1}{2}[\partial_t+\partial_t']\Psi_a&=&-\frac{a_t}{2a}\gamma^0\gamma^i[D_i+D_i']\Psi_a+\frac{ra_t}{2a}\gamma^0D_i'D_i\Psi_a-\beta_Y\Phi_{ab}\gamma^0(\chi,\xi)^T_b,\\
\frac{1}{2}[\partial_t+\partial_t'](\chi,\xi)^T&=&-\frac{a_t}{2a}\gamma^0\gamma^i[\partial_i+\partial_i'](\chi,\xi)^T+\frac{ra_t}{2a}\gamma^0D_i'D_i(\chi,\xi)^T-\beta_Y\Phi^\dagger\gamma^0\Psi.
\ea

The currents that are particularly relevant to us are the baryon and lepton currents which, when converted to male and female fermions, becomes
\ba
j^0_{(5)}&=&-\frac{i}{4}\left[-\bar\Psi_M(x)\gamma^0\gamma^5\Psi_F(x+0)-\bar\Psi_M(x+0)\gamma^0\gamma^5\Psi_F(x)\right.\\\nonumber
 &~&\qquad+\bar\chi_M(x)\gamma^0\gamma^5\chi_F(x+0)+\bar\chi_M(x+0)\gamma^0\gamma^5\chi_F(x)\\\nonumber
 &~&\qquad+\bar\xi_M(x)\gamma^0\gamma^5\xi_F(x+0)+\bar\xi_M(x+0)\gamma^0\gamma^5\xi_F(x)\\\nonumber
 &~&\qquad-\bar\Psi_F(x)\gamma^0\gamma^5\Psi_M(x+0)-\bar\Psi_F(x+0)\gamma^0\gamma^5\Psi_M(x)\\\nonumber
 &~&\qquad+\bar\chi_F(x)\gamma^0\gamma^5\chi_M(x+0)+\bar\chi_F(x+0)\gamma^0\gamma^5\chi_M(x)\\\nonumber
 &~&\qquad\left.+\bar\xi_F(x)\gamma^0\gamma^5\xi_M(x+0)+\bar\xi_F(x+0)\gamma^0\gamma^5\xi_M(x)\right],\\
j^k_{(5)}&=&-\frac{i}{4}\frac{a_t}{a}\left[-\bar\Psi_M(x)\gamma^k\gamma^5U_k(x)\Psi_F(x+k)-\bar\Psi_M(x+k)\gamma^k\gamma^5U_k^\dagger(x)\Psi_F(x)\right.\\\nonumber
 &~&\qquad\qquad+\bar\chi_M(x)\gamma^k\gamma^5\chi_F(x+k)+\bar\chi_M(x+k)\gamma^k\gamma^5\chi_F(x)\\\nonumber
 &~&\qquad\qquad+\bar\xi_M(x)\gamma^k\gamma^5\xi_F(x+k)+\bar\xi_M(x+k)\gamma^k\gamma^5\xi_F(x)\\\nonumber
 &~&\qquad\qquad-\bar\Psi_F(x)\gamma^k\gamma^5U_k(x)\Psi_M(x+k)-\bar\Psi_F(x+k)\gamma^k\gamma^5U_k^\dagger(x)\Psi_M(x)\\\nonumber
 &~&\qquad\qquad+\bar\chi_F(x)\gamma^k\gamma^5\chi_M(x+k)+\bar\chi_F(x+k)\gamma^k\gamma^5\chi_M(x)\\\nonumber
 &~&\left.\qquad\qquad+\bar\xi_F(x)\gamma^k\gamma^5\xi_M(x+k)+\bar\xi_F(x+k)\gamma^k\gamma^5\xi_M(x)\right],
\ea

\section{Fermion initial conditions}
\label{app:ferminit}

In order to specify the initial conditions we need some mode functions for the fermion fields. The ones we use are those that are vacuum fluctuations for a Higgs field taking the value $\phi=(0,v/\sqrt 2)$, $\tilde\phi=(v/\sqrt 2,0)$. The lattice equations of motion then read
\ba
\left[-\frac{1}{2}\gamma^0(\partial_0+\partial_0')-\frac{a_t}{2a}\gamma^i(\partial_i+\partial_i')+\frac{r\,a_t}{2a}\partial_{i}\partial_i'\right]
\left(\begin{array}{c}\chi\\ \xi\end{array}\right)
=\frac{va_t}{\sqrt{2}}\tilde{M}\Psi,\\
\left[-\frac{1}{2}\gamma^0(\partial_0+\partial_0')-\frac{a_t}{2a}\gamma^i(\partial_i+\partial_i')+\frac{r\,a_t}{2a}\partial_{i}\partial_i'\right]\Psi=\frac{va_t}{\sqrt{2}}M\left(\begin{array}{c}\chi\\ \xi\end{array}\right),\\
\ea
with $M$ and $\tilde{M}$ some general Yukawa mixing matrices.
This is a set of linear equations, so we perform a Fourier decomposition and look at positive frequency modes
\ba
\psi^+&=&U(k)e^{ik.x},
\ea
and find that by defining
\ba
s_0 = \sin(k_0\, a_t)/a_t=-\sin(\omega\, a_t)/a_t,\qquad s_i(k)=\sin(k_i)/a,\qquad m(k)=r\sum_i(1-\cos(k_i)),\nonumber\\
\ea
one gets
\ba
-i\gamma^\mu s_\mu U^{(\Psi)}&=&\frac{m(k)}{a}U^{(\Psi)}+\frac{v}{\sqrt 2}M
\left(
\begin{array}{c}
U^{(\chi)} \\
U^{(\xi)}
\end{array}
\right),\\
-i\gamma^\mu s_\mu
\left(
\begin{array}{c}
U^{(\chi)} \\
U^{(\xi)}
\end{array}
\right)
&=&\frac{m(k)}{a}
\left(
\begin{array}{c}
U^{(\chi)} \\
U^{(\xi)}
\end{array}
\right)
+\frac{v}{\sqrt 2}\tilde M\Psi,
\ea
now define $\tilde m=m(k)/a$ and $\tilde v=v/\sqrt 2$ and make the simplification that the Yukawa couplings are related by
\ba
G^{e\star}=G^u,\qquad G^{\nu\star}=-G^d,
\ea
so that defining
\ba
g^u=\sqrt{\tilde vG^u},\;g^d=\sqrt{\tilde vG^d},
\ea
allows us to diagonalize the equations by introducing the canonically normalized Fermi fields $\Omega_i$,
\ba
\Omega_1&=&\frac{1}{\sqrt 2|g^u|}[g^{u\star}\Psi_1+g^u\chi],\\
\Omega_2&=&\frac{1}{\sqrt 2|g^u|}[-g^{u\star}\Psi_1+g^u\chi],\\
\Omega_3&=&\frac{1}{\sqrt 2|g^d|}[g^{d\star}\Psi_2+g^d\xi],\\
\Omega_4&=&\frac{1}{\sqrt 2|g^d|}[-g^{d\star}\Psi_2+g^d\xi].
\ea
The positive frequency modes of these Fermi fields then satisfy
\ba
-i\gamma^\mu s_\mu U_1&=&(\tilde m+|g^u|^2)U_1,\\
-i\gamma^\mu s_\mu U_2&=&(\tilde m-|g^u|^2)U_2,\\
-i\gamma^\mu s_\mu U_3&=&(\tilde m+|g^d|^2)U_3,\\
-i\gamma^\mu s_\mu U_4&=&(\tilde m-|g^d|^2)U_4.
\ea
They are now in the usual form, and we can solve for the canonical mode functions, where in general
\ba
\left(-i\gamma^\mu s_\mu-\mu\right)U(k)&=&0,\\
\left(i\gamma^\mu s_\mu-\mu\right)V(k)&=&0,\\
s^0&=&+\sqrt{\mu^2+\underline s^2},
\ea
which are
\ba
\label{eq:modesUV}
U_s(\underline k)&=&\left(
\begin{array}{c}
-\sqrt{s^0(k)-\underline\sigma.\underline s}\;\eta_s \\
\sqrt{s^0(k)+\underline\sigma.\underline s}\;\eta_s 
\end{array}
\right),\qquad
V_s(\underline k)=\left(
\begin{array}{c}
\sqrt{s^0(k)-\underline\sigma.\underline s}\;\eta_s \\
\sqrt{s^0(k)+\underline\sigma.\underline s}\;\eta_s 
\end{array}
\right),
\ea
and $\eta_s=\left(\begin{array}{c}1\\0\end{array}\right),\;\left(\begin{array}{c}0\\1\end{array}\right)$. 

\section{Handmade Sphalerons}
\label{app:HMsphalerons}

The basic idea \cite{Manton:1983nd} is that the vacuum manifold of the Higgs scalar is a three-sphere, $S^3_{vac}$, and one wants to construct some other three-sphere in order to have non-trivial maps between them. At spatial infinity the Higgs field is in the vacuum, but spatial infinity is only a two-sphere, $S^2_\infty$, and the maps $S^2_\infty\rightarrow S^3_{vac}$ are topologically trivial. However we can construct a three-sphere by considering a loop in configuration space, $S^1_{loop}$, and forming the smash product $S^2_\infty\wedge S^1_{loop}$, which {\emph is} a three-sphere, and so one may construct non-trivial maps. This loop in configuration space means we have a family of configurations parametrized by an angle.

In practise this may done by parametrizing the $S^2_\infty\wedge S^1_{loop}$ three-sphere using the unit four-vector
\ba\nonumber
p_a&=&(\sin\Gamma\sin\theta\cos\varphi,\sin\Gamma\sin\theta\sin\varphi,\cos^2\Gamma+\sin^2\Gamma\cos\theta,\sin\Gamma\cos\Gamma(\cos\theta-1)),
\ea
which is an explicit representation of the smash product, where $\Gamma$ parametrizes $S^1_{loop}$, and $\theta,\;\varphi$ parametrize $S^2_\infty$. \footnote{Note that $0<\Gamma<\pi$, $0<\varphi<2\pi$, $0<\theta<\pi$ and that the circle at $\theta=0$ is mapped to the point (0,0,1,0), and the two-sphere at $\Gamma=0$ is mapped to the same point (0,0,1,0), thus confirming the smash product nature of the parametrization.}

Now we work with the usual complex doublet and write the set of configurations at infinity as
\ba
\phi_\infty&=&\frac{v}{\sqrt 2}
\left(
\begin{array}{c}
\sin\Gamma\sin\theta e^{i\varphi}\\
e^{-i\Gamma}(\cos\Gamma+i\sin\Gamma\cos\theta)
\end{array}
\right)
=
\left(
\begin{array}{c}
\phi_{1\infty}\\
\phi_{2\infty}
\end{array}
\right),
\ea
which leads to the ansatz for the full radial dependence of the family of configurations
\ba\nonumber
\phi_\Gamma(r,\theta,\varphi)&=&h(r)\phi_\infty(\Gamma,\theta,\varphi)+\frac{v}{\sqrt 2}[1-h(r)]
\left(
\begin{array}{c}
0 \\
e^{-i\mu}\cos\Gamma
\end{array}
\right),
\ea
where the radial profile function is a smooth monotonic function obeying $h(0)=0,\; h(\infty)=1$. This leads to
\ba
\phi_{\Gamma=0,\;\pi}(r,\theta,\varphi)&=&\frac{v}{\sqrt 2}
\left(
\begin{array}{c}
0\\
1
\end{array}
\right),\\\quad
\phi_{\mu=\pi/2}(r,\theta,\varphi)&=&\frac{v}{\sqrt 2}h(r)
\left(
\begin{array}{c}
\sin\theta e^{i\varphi} \\
\cos\theta
\end{array}
\right),
\ea
so we see that the start and end points of the $\Gamma$ loop just correspond to the vacuum, with the centre of the loop ($\Gamma=\pi/2$) being the sphaleron \cite{Manton:1983nd}.

Of course, we also need an ansatz for the gauge field, and this is achieved by constructing the unitary matrix
\ba
\Omega_\infty(\Gamma,\theta,\varphi)&=&
\frac{\sqrt 2}{v}
\left(
\begin{array}{cc}
\phi_{2\infty}^\star & \phi_{1\infty} \\
-\phi_{1\infty}^\star & \phi_{2\infty}
\end{array}
\right).
\ea
The gauge field is in the vacuum at infinity, and so we take the asymptotic ansatz to be
\ba
W_{j(\infty)}(\Gamma,\theta,\varphi)&=&-\frac{i}{g}\del_i\Omega_\infty(\Gamma,\theta,\varphi)\Omega^\dagger_\infty(\Gamma,\theta,\varphi),
\ea
with the full radial dependence of the family of gauge fields being given by
\ba
W_j(\Gamma,r,\theta,\varphi)&=&f(r)W_{j(\infty)}(\Gamma,r,\theta,\varphi),
\ea
where $f(r)$ is a smooth monotonic function satisfying $f(0)=0,\;f(\infty)=1$.


\end{document}